\documentclass{article}
\usepackage[utf8]{inputenc}
\usepackage{amsmath}
\usepackage{natbib}
\usepackage{bm}
\usepackage{graphicx}
\usepackage{listings}
\usepackage{longtable}
\usepackage{indentfirst}
\usepackage{caption}
\usepackage{subcaption}
\usepackage{setspace} 
\usepackage[left=2cm, right=2cm, top=2.00cm, bottom=2.00cm]{geometry}
\title{Stochastic volatility models with skewness selection}
\author{Igor Martins\footnote{Insper. igorfbm@al.insper.edu.br} and Hedibert Freitas Lopes \footnote{Insper. hedibertfl@insper.edu.br}}
\date{November 2023}

\begin{document}

\maketitle

\section*{Abstract}

\noindent This paper expands traditional stochastic volatility models by allowing for time-varying skewness without imposing it. While dynamic asymmetry may capture the likely direction of future asset returns, it comes at the risk of leading to overparameterization. Our proposed approach mitigates this concern by leveraging sparsity-inducing priors to automatically selects the skewness parameter as being dynamic, static or zero in a data-driven framework. We consider two empirical applications. First, in a bond yield application, dynamic skewness captures interest rate cycles of monetary easing and tightening being partially explained by central banks' mandates. In an currency modeling framework, our model indicates no skewness in the carry factor after accounting for stochastic volatility which supports the idea of carry crashes being the result of volatility surges instead of dynamic skewness.

\vspace{0.5cm}

\noindent \textbf{Key words:} Stochastic Volatility; Sparsity; Skewness.

\section{Introduction}

Accurate representation of asset returns is one of the key topics in finance. Based on the theoretical results from \cite{markowitz1952} and \cite{hansen1987role}, standard approaches to asset pricing have largely focused on the first and second moments. Stochastic volatility (SV) models discussed, for example, on \cite{jacquier2002bayesian}, are one of the cornerstone models in modern financial econometrics. In its simplest form SV models asset returns via normal distribution with persistent volatility and a mean that is either constant or a linear function of explanatory variables. Such models capture the first two moments of asset returns in a simple and elegant matter being supported empirically and theoretically as discussed in \cite{shephard2005stochastic}. 

While we acknowledge the importance of the first two moments in asset pricing, we also notice the potential benefits of including skewness when modeling returns. Due to its ability of capturing the likely direction of returns, models with time-varying skewness may be more suitable to forecast periods with a higher concentration of same signal returns leading to better detection of periods of both overperformance and underperformance. \cite{bianchi2022taming} is one key example of the empirical benefits of adding such feature for modeling cross-sectional stock momentum. While the momentum factor is known for delivering good mean-variance compensation, it is also subject to long period of negative performance. By capturing such prolonged periods of likely negative returns via dynamic skewness, \cite{bianchi2022taming} improves the performance of the stock momentum factor upon traditional approaches which neglect skewness. 

However, including dynamic skewness in traditional financial econometric models is not a free-lunch. While allowing for asymmetry may lead to a better representation of some financial time series, it may not be a vital feature and its inclusion risks overparameterization. Therefore, we wish to include dynamic skewness only when required by the data and remove such feature if is not necessary. 

This paper expand stochastic volatility models by allowing dynamic skewness without having to impose it. We replace the traditional hypothesis of gaussian errors in favor of a skew-normal distribution. Such change preserves the usual features for the first two moments of SV models but allows for dynamic skewness. Since the inclusion of time-varying asymmetry may not always be necessary, we consider a sparsity - inducing scheme for its parameters. In particular, we consider a random-walk evolution for the asymmetry. When the standard deviation of the dynamic process is shrunk to zero, our model results in a SV with constant skewness. If, additionally, the level of the skewness is shrunk to zero, we recover a traditional SV model. By combining prior information with the likelihood of the model, our proposed approach automatically chooses between dynamic, static or no skewness.  

We consider two empirical applications. We obtain three main results on a bond yield application for Brazil and the US. First, our proposed model indicates that bond yield changes for both countries are better represented by including time-varying skewness with out-of-sample improvements for forecasting the direction of future yields. Second, the recovered skewness is associated with interest rate cycles of monetary easing and tightening. Third, inflation and unemployment partially explain the recovered skewness linking it to central banks' mandates. We also model the carry factor for currency returns. Our model indicates not only the lack of dynamics but also no skewness at all for it after accounting for volatility. Therefore, similar to the crash mitigation via volatility scaling proposed by \cite{barroso2015momentum}, the carry factor also have its crashes mitigated once dynamic volatility is taken into account without requiring the inclusion skewness.

This paper intersects and contributes to multiple areas. First, it contributes to the stochastic volatility literature by expanding the static skewness model of \cite{nakajima2012stochastic} to the dynamic case while also expanding the sparsity - inducing scheme of \cite{nakajima2020skew} by allowing dynamic skewness to be shrunk towards the static case. Second, it contributes to the toolbox of methods for recovering dynamic skewness. \cite{trolle2014swaption} rely on option data while \cite{rafferty2012currency} uses rolling windows. Both approaches have limitations. While theoretically sound, option based approaches require tradeable options which large collection of strikes with high liquidity and continuous expiration dates. Such requirements are hard to meet in several applications, specially when for emerging markets such as in our Brazilian bond applications. Rolling-window based approaches artificially input dynamics into skewness by changing the sample period by period. Such change comes at the cost of outliers playing a large role in the estimation in addition to a trade-off based on window-size, which will affect the precision of the estimate, and the speed in which the dynamics change. Third, it contributes to the interest rate literature by showing that inflation and unemployment partially explain skewness and linking it to central banks' mandates expanding the traditional mean and variance analysis of papers such as \cite{litterman1991common} and \cite{joslin2018can}. Fourthly, it contributes to the debate of whether the carry factor presents dynamic skewness after accounting for heteroskedasticity, discussed \cite{burnside2011peso} and \cite{jurek2014crash}, by claiming that skewness is unlikely to be dynamic and, in fact, is more likely to be zero after accounting for SV effects. 

The paper is organized as follows. It starts by describing traditional SV models and moves on to our proposal with time-varying skewness on Section \ref{Sec:Model}. Section\ref{Sec:Sparsity} discusses the sparsity-inducing framework. Section \ref{Sec:Posterior} shows our HMC approach to simulate from the joint posterior. Section \ref{Sec:BondYields} presents the bond yield forecasting application while Section \ref{Sec:Carry} shows the carry factor application. Section \ref{Sec:Conclusion} concludes. 
 
\section{SV model with time-varying skewness} \label{Sec:Model}

A vanilla stochastic volatility model is given by Equations (\ref{SV_vanilla_obs_eq}) - (\ref{SV_vanilla_error_latent}). Equation (\ref{SV_vanilla_obs_eq}) represent asset returns with mean 0 and dynamic volatility $exp(h_t/2)$. Equation (\ref{SV_vanilla_latent_eq}) indicates 
persistent log-volatility with level $\mu$ and persistence $\phi$ with initial values given by the stationary distribution shown in Equation (\ref{SV_vanilla_init_dist}). In its simplest version both the measure and state equations have Gaussian errors as represented in Equations (\ref{SV_vanilla_error_obs}) and (\ref{SV_vanilla_error_latent}). 

\begin{equation}\label{SV_vanilla_obs_eq}
    y_t = exp\Bigg(\frac{h_t}{2} \Bigg) \varepsilon_t  
\end{equation}
\begin{equation}\label{SV_vanilla_latent_eq}
    h_{t} = \mu_h + \phi_h (h_{t-1} - \mu_h) + \sigma_h \eta_t 
\end{equation}
\begin{equation}\label{SV_vanilla_init_dist}
        h_0 \sim N \Bigg( \mu_h, \frac{\sigma^2_h}{1 - \phi_h^2} \Bigg)
\end{equation} 
\begin{equation} \label{SV_vanilla_error_obs}
    \varepsilon_t \sim N(0,1)
\end{equation}
\begin{equation} \label{SV_vanilla_error_latent}
    \eta_t \sim N(0,1)
\end{equation}

\begin{figure}[h!]
\centering
\includegraphics[height= 6cm,width=0.95\textwidth]{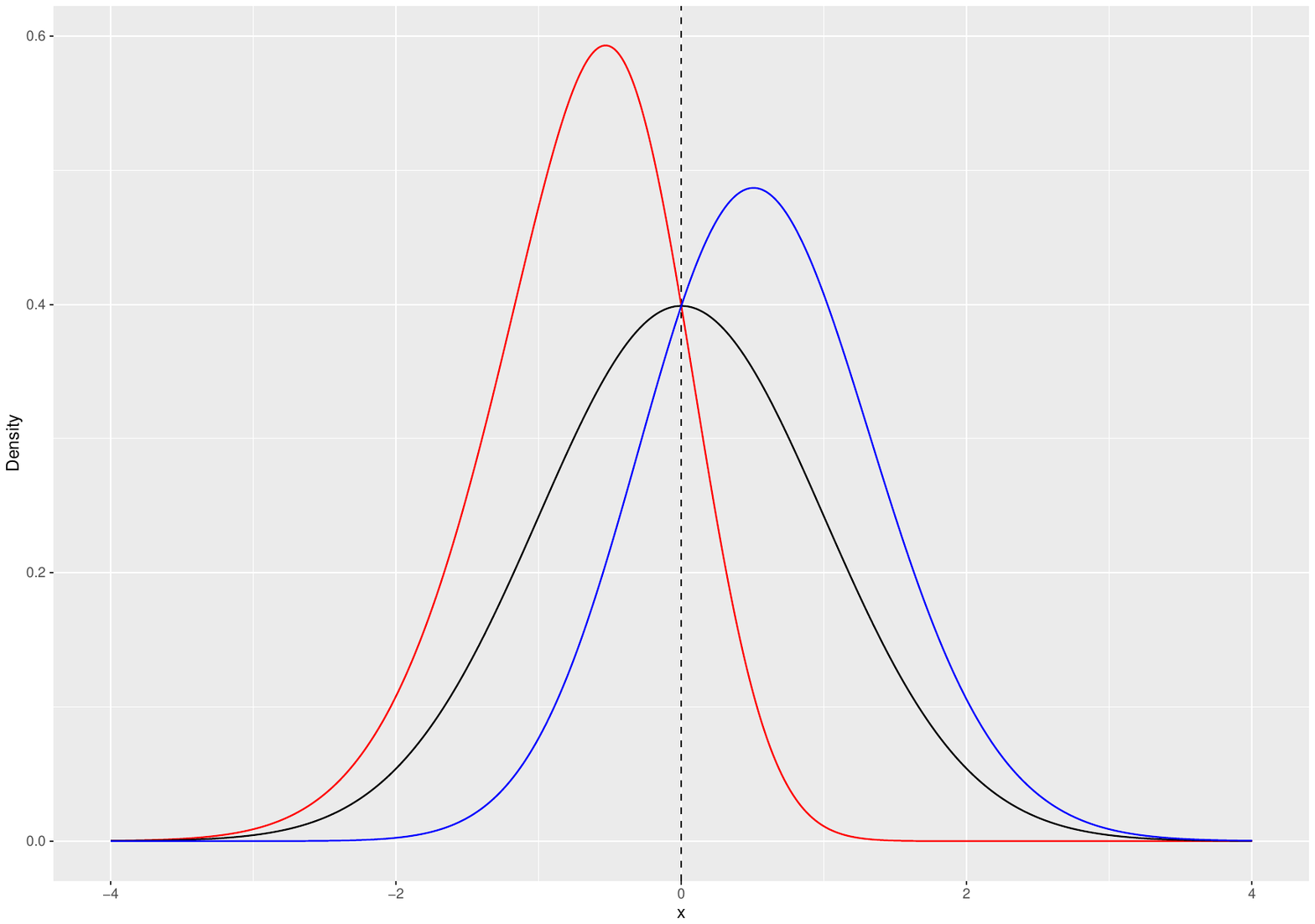}
\caption{Skew-normal densities with $\xi = 0$ and $\omega =1$ and $\lambda$ equal to -2, 0, and 1 for the red, black and blue lines, respectively. If $\lambda < 0$, the distribution is skewed to the left. If $\lambda = 0$, SN(0) reduces to the standard Gaussian distribution. Finally, if $\lambda > 0$ the distribution is skewed to the right. } 
\label{Fig:NS_multiple_lambda}
\end{figure}

In order to introduce asymmetry into the model, we replace the normal variable $\varepsilon_t$ in Equation (\ref{SV_vanilla_error_obs}) by a skew-normal random variable $z_t$. We consider the skew-normal representation of \cite{azzalini1985class} and \cite{azzalini2013skew} shown in Equation (\ref{Eq:Std_SN}) where $\phi (\cdot)$ and $\Phi (\cdot)$ are the standard normal density and distribution function, respectively. $\lambda$ controls the degree of asymmetry in the distribution as show in Figure (\ref{Fig:NS_multiple_lambda}). In particular, for $\lambda = 0$, the skew-normal reduces to a standard normal distribution. Also, we may add location $\xi$ and scale $\omega$ parameters to the skew-normal denoting  $X = \xi + \omega z$ by $ X \sim SN(\xi, \omega, \lambda)$ which has its distribution presented in Equation (\ref{Eq:SN_with_location_scale}). Appendix A describes how $\xi$ , $\lambda$ and $\lambda$ relate to mean, variance and skewness. Therefore, our model can be viewed as an expansion of the stochastic volatility models with static skewness proposed by \cite{nakajima2012stochastic}, \cite{nakajima2020skew} and \cite{li2020leverage}. 

\begin{equation}\label{Eq:Std_SN}
    \text{If Z} \sim SN(\lambda) \text{, then } p(z;\lambda) = 2 \phi(z) \Phi(\lambda z)
\end{equation}

\begin{equation}\label{Eq:SN_with_location_scale}
    \text{If X} \sim SN(\xi, \omega, \lambda) \text{, then } p(x; \xi, \omega, \lambda) = 2 \frac{1}{\omega} \phi \Big( \frac{x - \xi}{\omega} \Big) \Phi \Big(\lambda \frac{x - \xi}{\omega}  \Big)
\end{equation}

As shown in \cite{bianchi2022taming} and \cite{carr2007stochastic}, its plausible that asset returns have time-varying skewness. Thus, we allow for this possibility by replacing the static $\lambda$ for $ \lambda_t$ which evolves according to a random-walk starting at $\lambda_0$ as represented by Equations (\ref{Eq:zt_dyn}) to (\ref{Eq:zt_rw}). Therefore, we change our observation equation by allowing both volatility and skewness to be time-varying changing the observation equation to Equation (\ref{SVSkew_obs_eq}).

\begin{equation}\label{Eq:zt_dyn}
   z_t \sim SN(\lambda_t) 
\end{equation}
\begin{equation}\label{Eq:zt_rw}
   \lambda_t = \lambda_{t-1}  + \sigma_{\lambda} \eta_{\lambda,t} \text{ with } \eta_{\lambda,t} \sim N(0,1)
\end{equation}
\begin{equation}\label{Eq:zt_init}
   \lambda_0 = \alpha_0
\end{equation}
\begin{equation}\label{SVSkew_obs_eq}
    y_t = exp\Bigg(\frac{h_t}{2} \Bigg) z_t  
\end{equation}

\section{Sparsity-inducing approach}\label{Sec:Sparsity}

The inclusion of asymmetry may lead to a better representation of time series specially by capturing periods with a higher mass of returns with the same signal, as shown, for example, in \cite{bianchi2022taming}, \cite{azzalini2013skew}, and \cite{rachev2005fat}. However, it may not always be necessary since it requires the estimation of additional parameters. To mitigate this overparametrization risk, we aim to include time-varying skewness only when required by the data. While one can estimate multiple models varying the skewness specification and then performing selection via Bayes factor, we propose a sparse-inducing method which conveniently performs model selection without requiring estimating multiples models while also avoiding Bayes factor completely. 

Equation (\ref{Eq:zt_rw}) drives the dynamics of the asymmetry parameter. If $\sigma_{\lambda}$ is close enough to zero, then $\lambda_t$ is static and assume the value of its initial point $\lambda_0$. Additionally, if the initial point is also zero, then the model reduces to the vanilla SV model. Therefore, if we induce sparsity for $\sigma_{\lambda}$ and $\alpha_0$ we can achieve all 3 cases of interest.

In the non-Bayesian literature, shrinkage is based on maximizing the likelihood of a model subject to a penalty function with the LASSO of \cite{tibshirani1996regression} being the most commonly employed approach. From a Bayesian point of view, shrinkage problems can be represented as a penalization to the log-likelihood via log-prior. In fact, the posterior mode of a linear model with Double Exponential prior with location 0 and scale $2/\psi$ is equal to the point estimate of the LASSO with penalty $\psi$ as shown in \cite{park2008bayesian}. 

Thus, in order to shrink $\sigma_{\lambda}$ and $\lambda_0$ towards zero, we consider both having Double Exponential priors shown in Equation (\ref{Eq:DE_prior_sigma}) and (\ref{Eq:DE_prior_alpha0}) similarly to the approach of \cite{belmonte2014hierarchical} in the context of dynamic regression models. Priors such as the ones discussed in \cite{lopes2022parsimony} and \cite{bitto2019achieving} also are reasonable approaches to induce sparsity on time-varying parameter models. 

\begin{equation}\label{Eq:DE_prior_sigma}
    \sigma_{\lambda} \sim \text{DoubleExponential}(0, 1/\kappa_{\sigma_{\lambda}}) \text{ with } \kappa_{\sigma_{\lambda}} \sim Gamma(a,b)
\end{equation}

\begin{equation}\label{Eq:DE_prior_alpha0}
    \alpha_{\lambda} \sim \text{DoubleExponential}(0, 1/\kappa_{\alpha}) \text{ with } \kappa_{\alpha} \sim Gamma(c,d)
\end{equation}

While, to the best of our knowledge, we are the first to induce sparsity on the dynamic skewness framework, \cite{nakajima2020skew} have used the spike and slab prior of \cite{george1993variable} to produce a data-driven framework which obtains the posterior probability of the presence of static skewness.  

\section{Remaining priors and posterior inference}\label{Sec:Posterior}

Our modeling approach leads to the following unknown quantities $ \Theta = \{ \mu_h, \phi_h, \sigma_h, \alpha_{0}, \sigma_{\lambda}, \{h_t\}, \{\lambda_t\} \}$. We aim to recover the joint posterior distribution $p(\Theta|y)$ which by Bayes' rule can be computed as follows
$$ p(\Theta|y) = \frac{p(y|\Theta) p(\Theta)}{\int p(y|\Theta) p(\Theta) d\Theta } $$
$p(y|\Theta)$ is the likelihood component being characterized by our proposed sampling model.  We consider independent components for each member of $p(\Theta)$. $p(\mu_h)$ and $p(\phi_h)$ follow Gaussian distributions, $p(\sigma_h^2)$ has inverse gamma distribution while $p(\alpha_0)$ and $p(\sigma_{\lambda})$ have double exponential distribution as discussed previously. The exact values for the prior parameters are described in Appendix B and Appendix C.  

While we can provide a full description of $p(y|\Theta)$ and $p(\Theta)$, the problem of recovering $p(\Theta|y)$ is not analytically tractable. Thus, we resort to simulation methods to sample from $p(\Theta|y)$ and then compute summary posterior statistics to characterize our results. This paper uses a particular Markov Chain Monte Carlo (MCMC) method known as Hamiltonian Monte Carlo (HMC) instead of the more traditional Random - Walk Metropolis-Hastings (RWMH) algorithm to sample from $p(\Theta|y)$. 

As discussed in \cite{gamerman2006markov}, the RWMH algorithm is a type o MCMC algorithm which generates a sequence of values $\Theta$ which approximates  $p(\Theta|y)$. Each iteration i generates $\Theta^{(i)}$ which is defined in part by $q(\Theta^{prop}|\Theta^{i-1})$ where $\Theta^{prop}$ is a proposal for the next value in the chain and $\Theta^{i-1}$ represents the value of $\Theta$ on the previous iteration. The name RWMH is due to the proposal being generate as a random - walk from the previously sampled $\Theta$. Each proposed value for $\Theta^{i}$ may be accepted or rejected based on Equation (\ref{Eq:AcceptanceRatio}) where acc represents the acceptance ratio. While there are little restrictions for the use of RWMH algorithms, its limitations come from the computational side since the acceptance ratio is usually low requiring a large amount of iterations.

\begin{equation}\label{Eq:AcceptanceRatio}
   acc = min \Bigg(1, \frac{f(\Theta^{prop}) q(\Theta^{t-1}|\Theta^{prop}) }{f(\Theta^{t-1}) q(\Theta^{prop}|\Theta^{t-1}) } \Bigg)  
\end{equation}

One of the main advantages of HMC comes exactly due to its higher acceptance rate than RWMH. HMC improves upon RWMH by employing guided proposals based on the gradient of the log posterior to direct the Markov chain towards regions of higher posterior density while also sampling the tail areas properly as discussed in \cite{betancourt2017conceptual}. \cite{thomas2021learning} and \cite{betancourt2017conceptual} are good introductions to HMC. 

Over time, HMC travels on trajectories that are governed by the Hamiltonian equations: 
$$ \frac{dp}{dt} = - \frac{\partial H(\Theta, p)}{\partial \Theta} = \nabla_{\Theta} log f(\Theta|y)$$
$$ \frac{d\Theta}{dt} = \frac{\partial H(\Theta, p)}{\partial p} = \frac{\partial K(p)}{\partial p} = M^{-1}p $$

where $H(\Theta, p)$ represents the Hamilton function. In the context of this paper $H(\Theta, p) = -log f(\Theta|y) + \frac{1}{2} p^{T} M^{-1} p$. Additionally, $\nabla_{\Theta} log f(\Theta|y)$ is the gradient of the log posterior density which is the main responsible for the improvement of HMC over RWMH in acceptance ratio.  
 
As discussed in \cite{betancourt2017conceptual}, proposals generated from the exact solution of the Hamilton equations would be accepted with probability 1. However, they usually are not analytically tractable and solutions comes from numerical methods, typically, via the leapfrog method. Since the leapfrog is an approximation, an acceptance step is added to ensure that proposals does not deviate far from the specified Hamiltonian $H(\theta, p)$. Therefore, on practice, the acceptance rate of HMC is less than 100\% but higher than RWMH. Thus, we employ HMC to sample from $p(\Theta|y)$. 

An additional benefit of HMC apporaches are their easy implementation via Stan introduce by \cite{carpenter2017stan}. In this paper, we combine Stan with R via rStan introduced by \cite{guo2020package}. In every application, we run our HMC scheme for 30000 iterations with the first 15000 being used as burn-in draws.

\section{Empirical application: Bond Yields}\label{Sec:BondYields}

The bond market is one of the largest in the world being key for investors and policy makers. Most papers focus on the first two moments e.g. \cite{litterman1991common}, \cite{collin2002bonds}, \cite{cochrane2005bond} and \cite{joslin2018can}. Our paper focus on the much less explored third moment. As discussed previously,  skewness captures the likely direction of returns. Therefore, it allows an interest rate investor to improve their forecast about sign of future yields allowing larger investments when positive returns are likely and buying protection or going short otherwise. 

\begin{figure}[h]
\centering
\begin{subfigure}[b]{0.95\textwidth}
\includegraphics[height=6.5cm,width=0.95\textwidth]{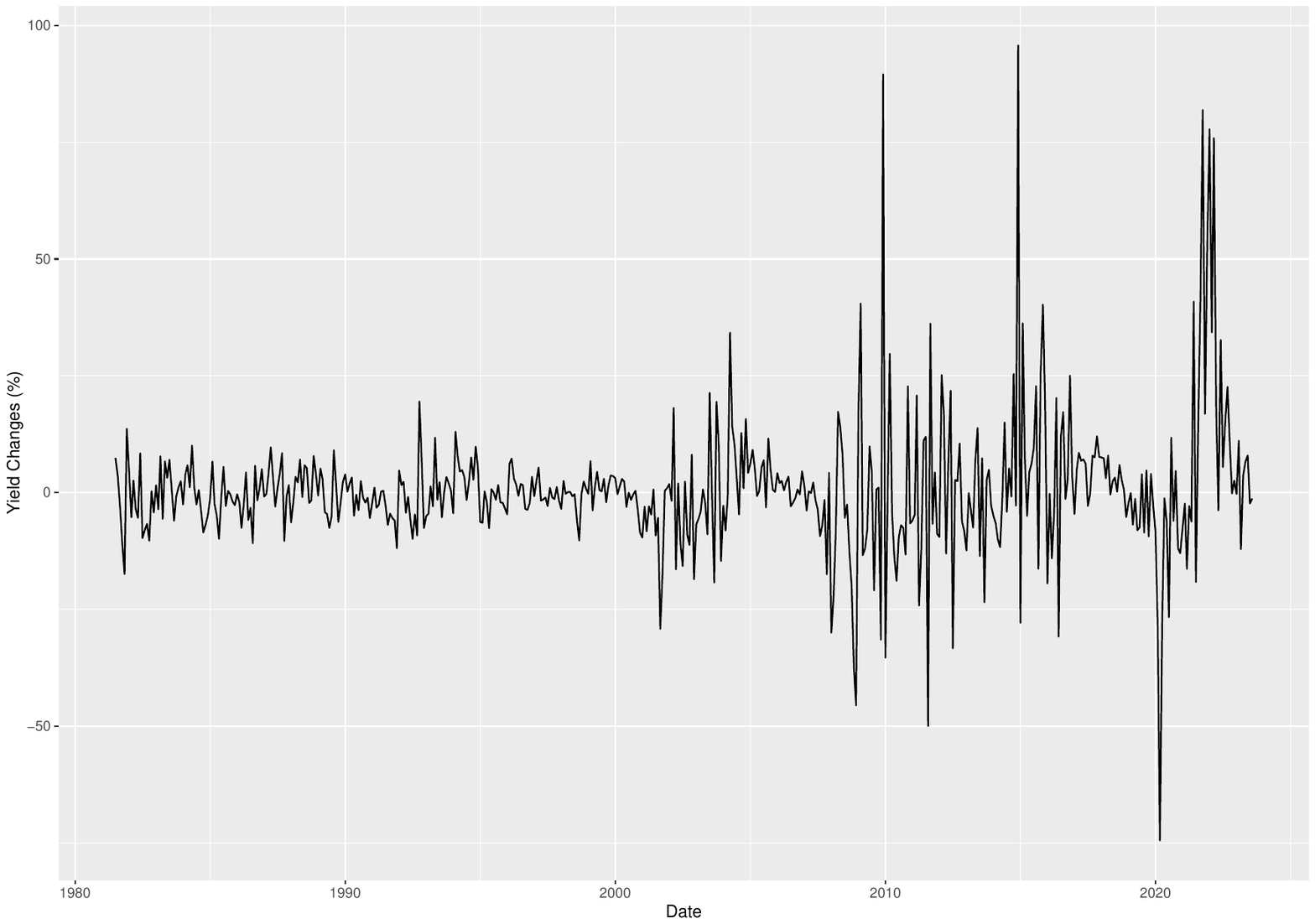} 
   \label{fig:us_yields} 
\end{subfigure}
\begin{subfigure}[b]{0.95\textwidth}
\includegraphics[height=6.5cm,width=0.95\textwidth]{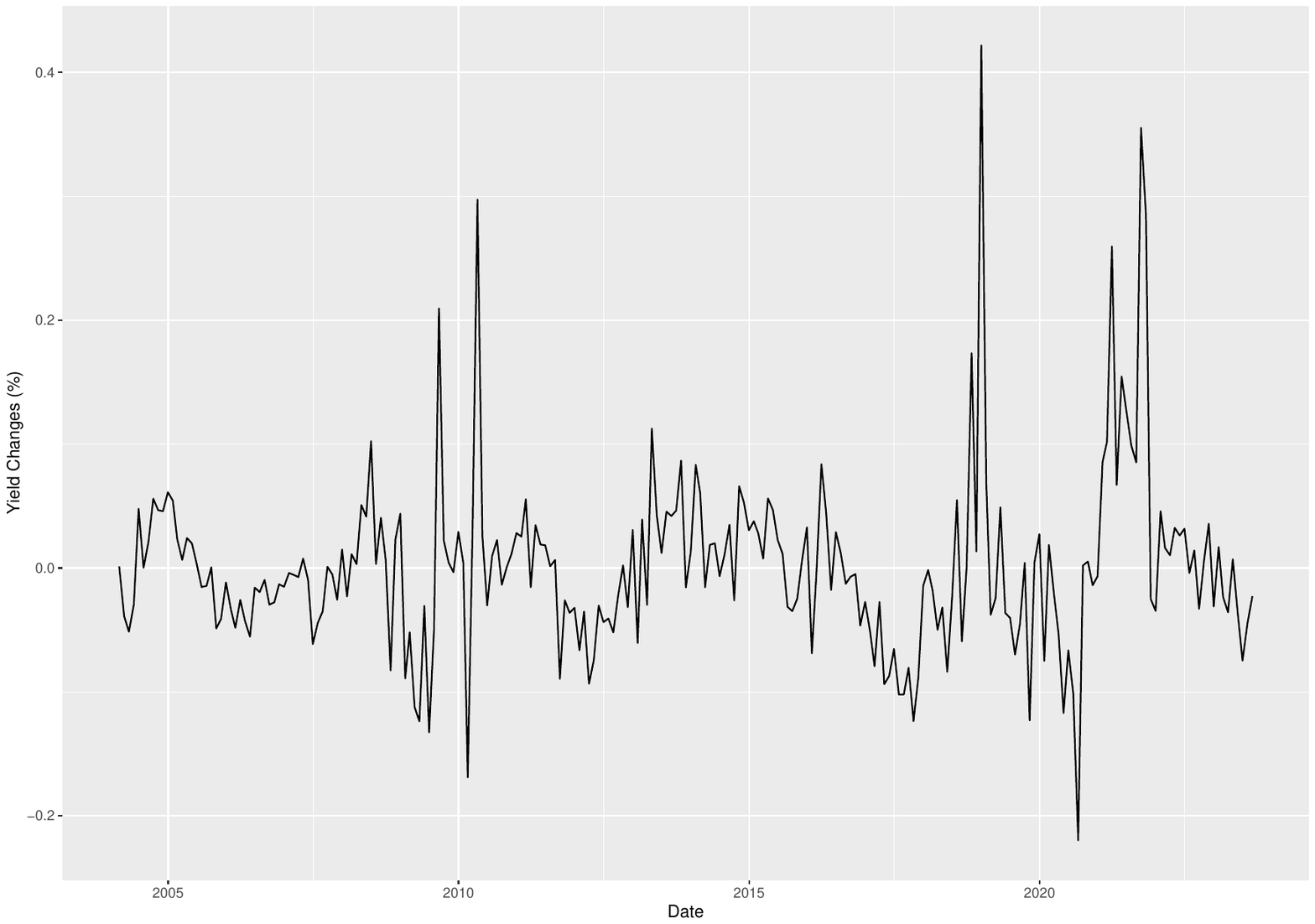} 
   \label{fig:br_yields}
\end{subfigure}
\caption{Change in yields from fixed 1-year maturity bonds from the US and Brazil. The top panel presents monthly changes in yields from 1-year bonds from the US government starting in 07/1981 and going up to 08/2023. Similarly, the bottom panel shows monthly changes in 1-year Brazilian bonds from 02/2004 up to 08/2023. }
\label{Fig:Yields_time_series}
\end{figure}

In our first application, we model monthly yield changes in fixed 1-year maturity for both American and Brazilian bonds. The sample of US bonds comes from the updated dataset of \cite{gurkaynak2007us} made available by the Federal Reserve Board starting at 07/1981 and going until 08/2023. The sample of Brazilian bonds is based on the DI interest rate contracts available on the Brazilian stock and futures exchange B3. The DI contracts are interpolated to form 1-year fixed maturity bonds using the same Nelson-Siegel procedure described in \cite{gurkaynak2007us} resulting in a sample starting in 02/2004 and ending in 08/2023. Figure (\ref{Fig:Yields_time_series}) plots both time series of monthly yield changes. Notably, both series fluctuate around zero with volatility clusters. Both series hint at the possibility of skewness. For example, the US series presents a persistent and negative yield change period in the early 90 and early 2000. In the Brazilian series, the negative and persistent yield changes are even clearer with the period from 08/2016 to 03/2018 being one example.  

\begin{figure}[h!]
\centering
\begin{subfigure}[b]{0.95\textwidth}
\includegraphics[height=5.5cm,width=0.95\textwidth]{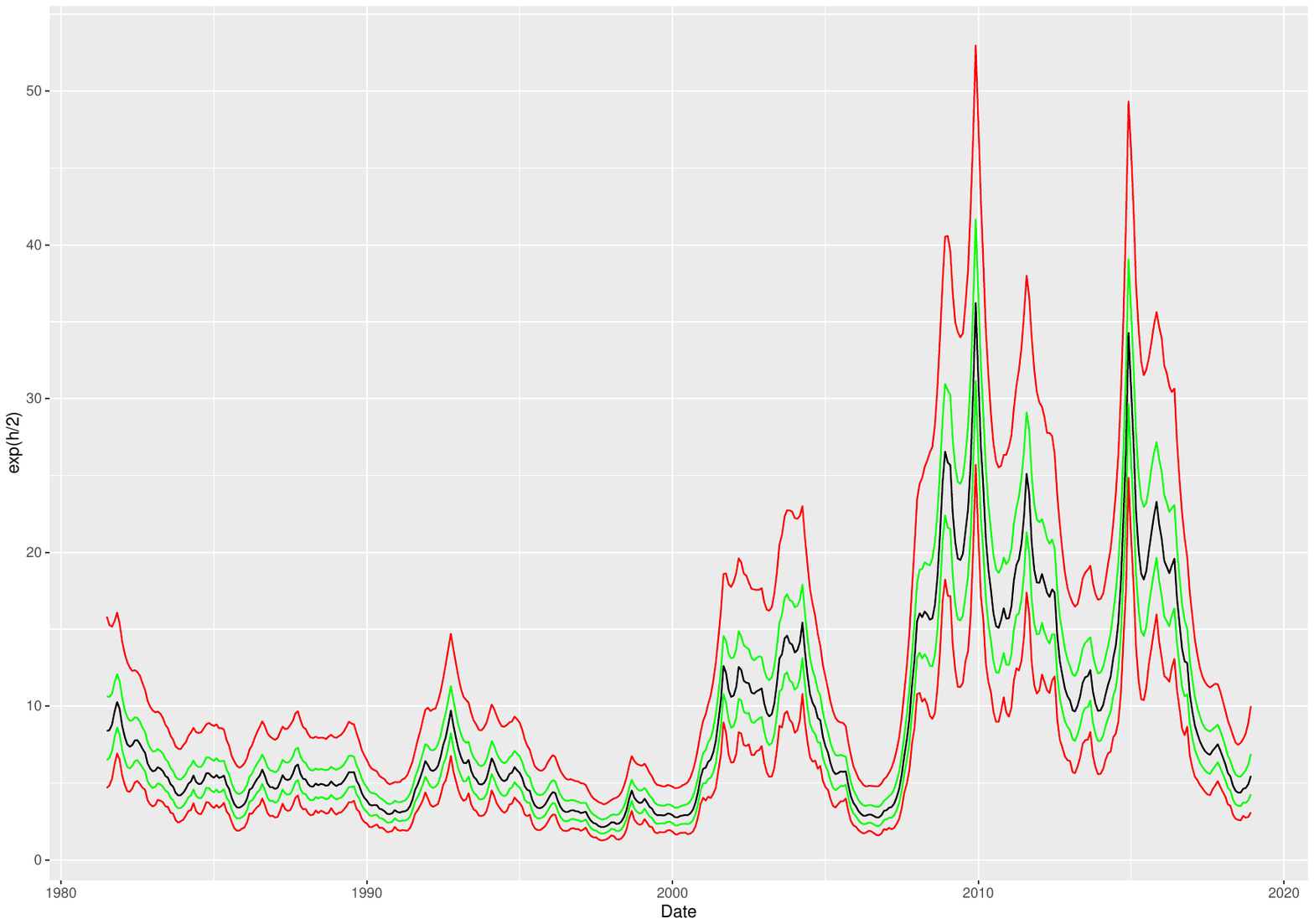} 
    \label{fig:us_scale} 
\end{subfigure}
\begin{subfigure}[b]{0.95\textwidth}
\includegraphics[height=5.5cm,width=0.95\textwidth]{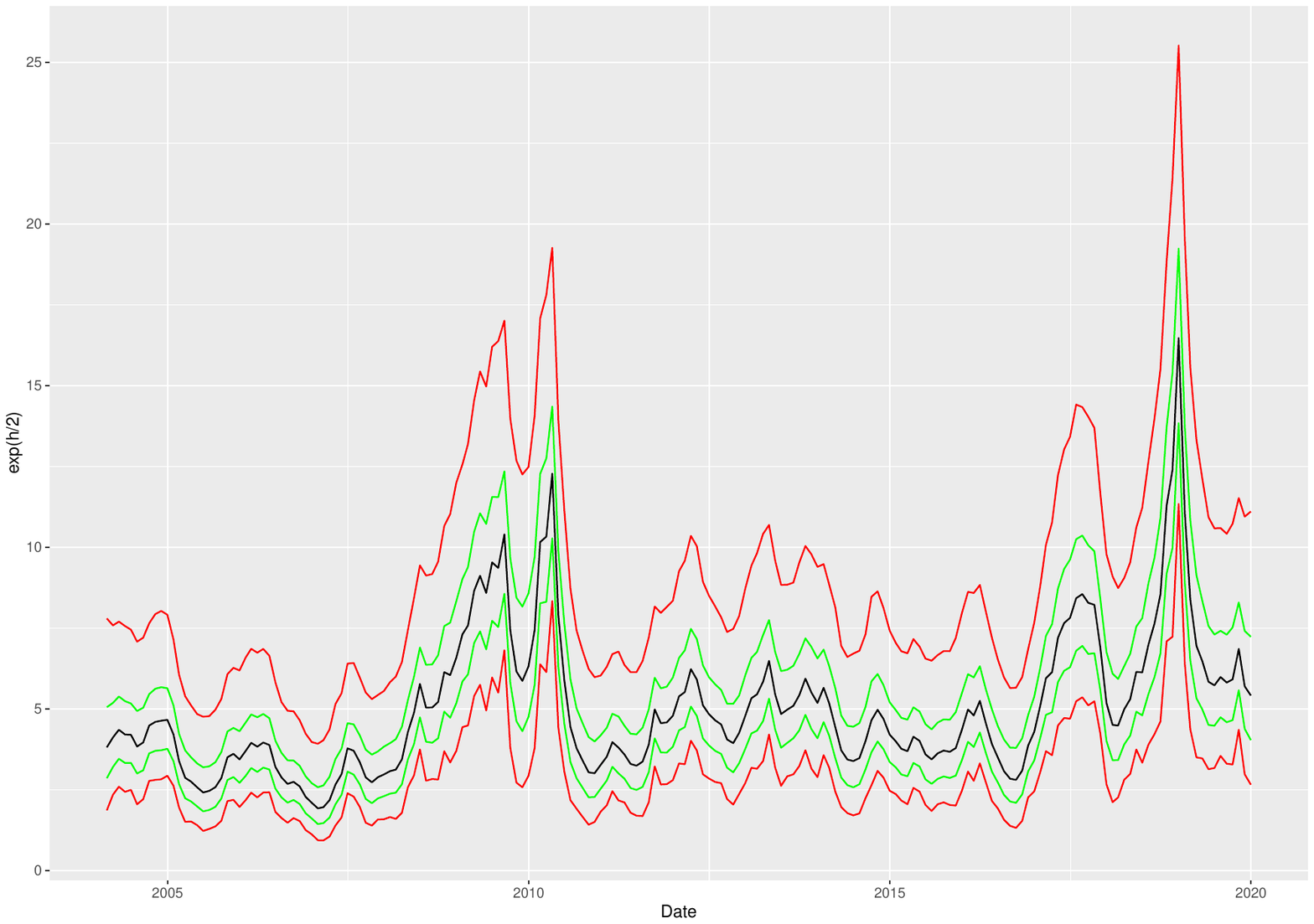} 
   \label{fig:br_scale}
\end{subfigure}
\caption{Scale recovered for the US and Brazilian bonds during the estimation period which goes up to 12/2018 for the US and up to 12/2019 for the Brazilian case. Green lines represent quantiles .25 and .75 while solid red lines represent quantiles .05 and .95. For both series, time-varying volatility is plausible.  Top panel indicates rises in yield change volatility during the early 00's and around the global financial crisis for the US. While the bottom panel indicate peaks of volatility for Brazilian yield changes between 2009 to 2010 with a second spike in late 2018 and early 2019. }
\label{Fig:Yield_Scale}
\end{figure}

We apply our proposed model to both US and Brazilian bonds with the priors presented in Appendix B. To account for the Brazilian time series being shorter than the American, the estimation sample comes closer to the present than the American. Precisely, we consider an estimation period which goes up to 12/2018 for the US and up to 12/2019 to the Brazilian case. Table (\ref{Tab:ParametersProposal}) presents the posterior summaries for the parameters of both time-series. The values of $\phi_h$ indicate high persistence in the log-scale which is reflected in Figure (\ref{Fig:Yield_Scale}) which plots posterior summaries of $\{h_t\}$. In both cases, our approach captures the surges in volatility indicated in Figure (\ref{Fig:Yields_time_series}). For example, the US series captures spikes of volatility in the early 2000's and on the financial crisis of 2008. The Brazilian time series also presents a spike in volatility in 2008 which is also captured by our model.

\begin{figure}[h!]
\centering
\begin{subfigure}[b]{0.95\textwidth}
\includegraphics[height=6cm,width=0.95\textwidth]{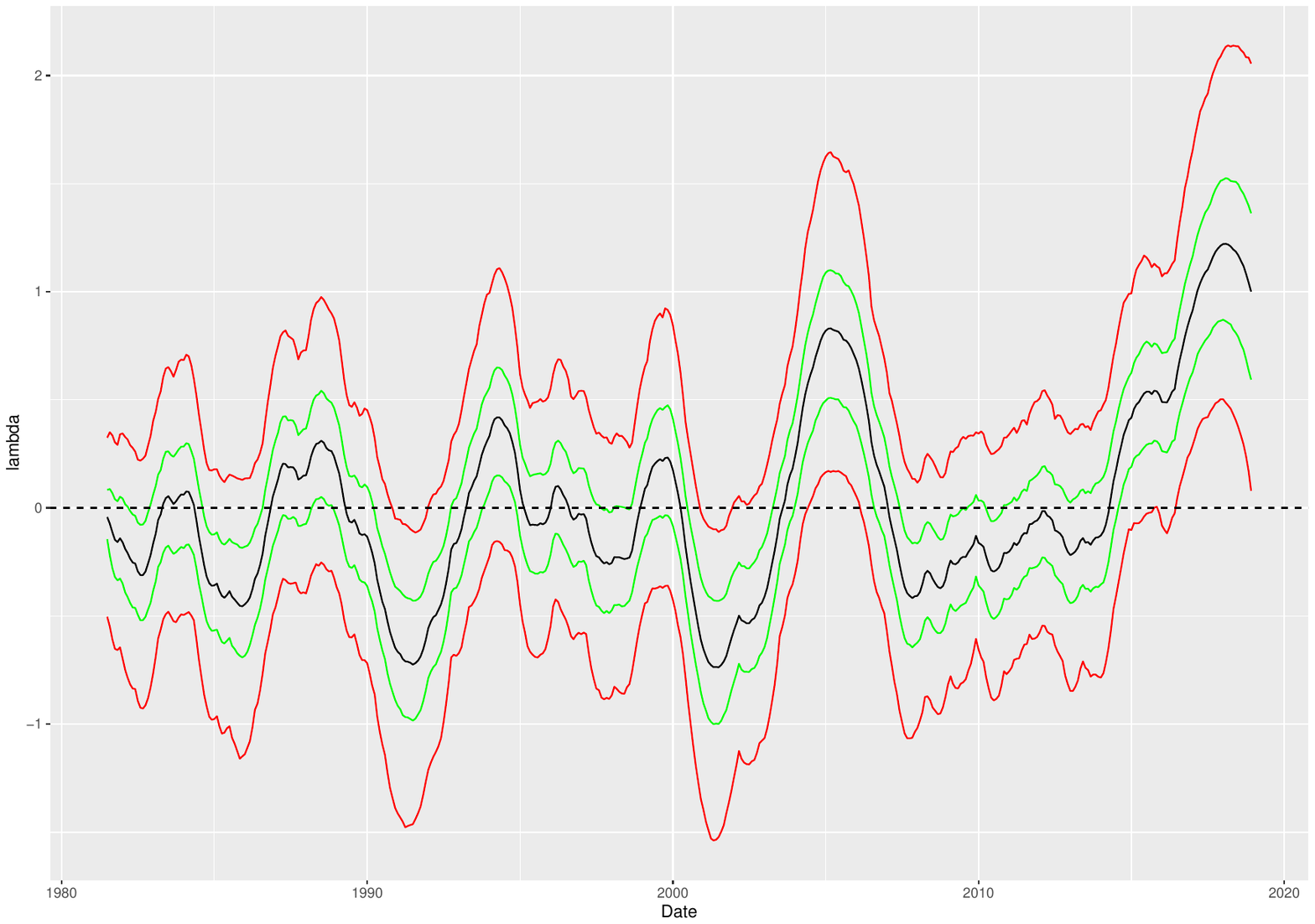} 
   \caption{}
   \label{fig:us_lambda} 
\end{subfigure}
\begin{subfigure}[b]{0.95\textwidth}
\includegraphics[height=6cm,width=0.95\textwidth]{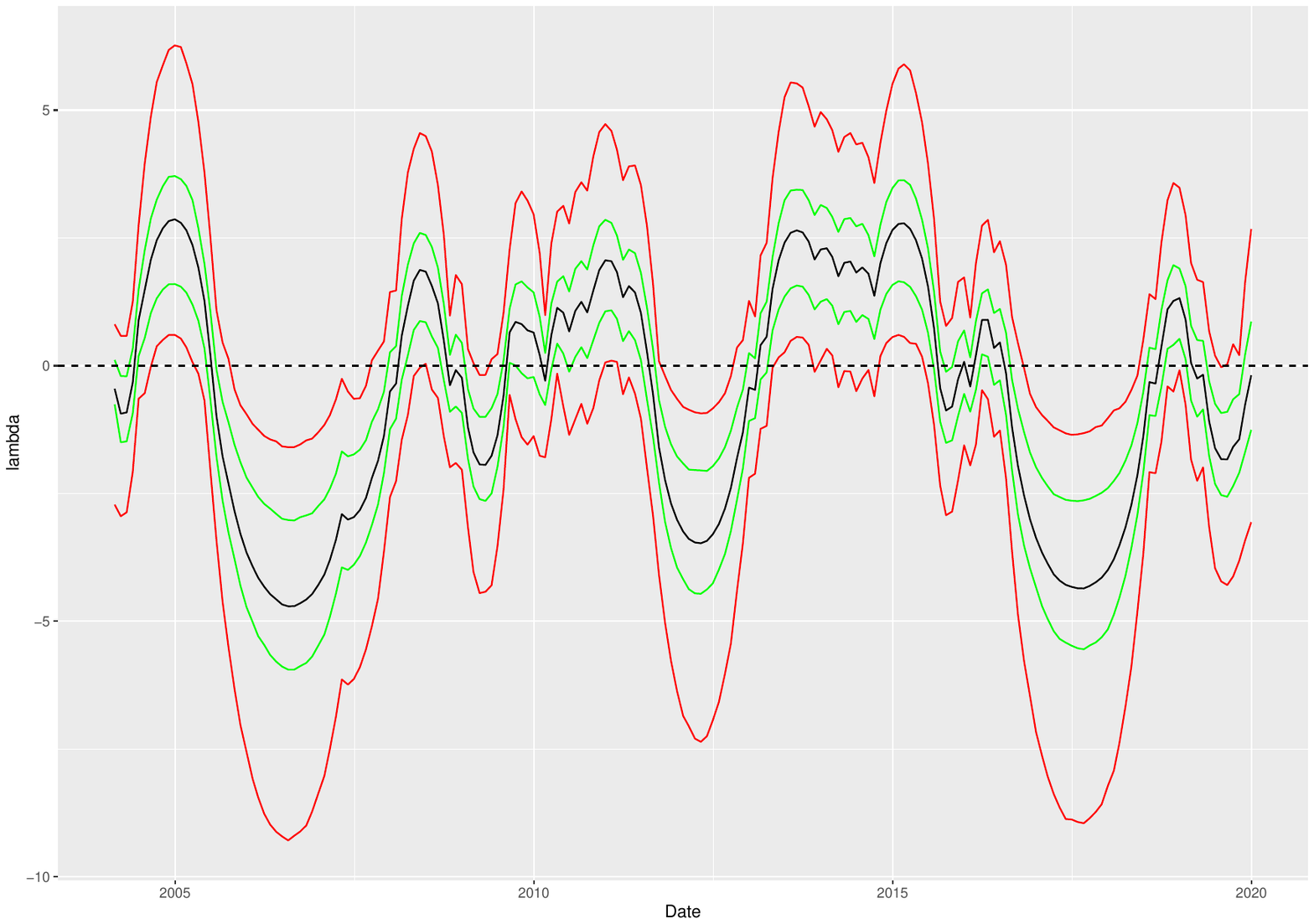} 
   \caption{}
   \label{fig:br_lambda}
\end{subfigure}
\caption{$\{\lambda_t\}$ recovered for the US and Brazilian bonds during the estimation period which goes up to 12/2018 for the US and up to 12/2019 for the Brazilian case. While $\lambda$ is not the skewness itself, it can be transformed by using the formula in Appendix A. Green lines represent quantiles .25 and .75 while solid red lines represent quantiles .05 and .95. Time-varying skewness is likely in both cases. We obtain stronger evidence o dynamic skewness for Brazil in the bottom panel than for the US in the top one.}
\label{Fig:lambda_plots}
\end{figure}

Additionally, both series are likely to have a dynamic skewness as show by the posterior summaries of $\sigma_{\lambda}$ with the Brazilian bonds having a bigger range of variation for $\lambda_t$ than bonds from the US. Such, evidence is supported by the posterior summaries of $\{\lambda_t\}$ in Figure (\ref{Fig:lambda_plots}) as well.  

\begin{table}[ht]
\centering
\begin{tabular}{lccccc}
  \hline
 & $\mu_h$ & $\phi_h$ & $\sigma_h$ & $\sigma_{\lambda}$ & $\alpha_{0}$ \\ 
   \hline
US q05 & 2.94 & 0.92 & 0.32 & 0.09 & -0.62 \\ 
US Mean  & 3.88 & 0.96 & 0.42 & 0.17 & -0.06 \\ 
US q95 & 4.84 & 0.99 & 0.55 & 0.27 & 0.42 \\ 
   \hline
BR q05 & 2.59 & 0.59 & 0.38 & 0.63 & -2.67 \\ 
BR Mean& 3.08 & 0.79 & 0.63 & 1.09 & -0.46 \\ 
BR q95 & 3.59 & 0.94 & 0.91 & 1.75 & 0.79 \\  
   \hline
\end{tabular}
\caption{Posterior summaries of the parameters from our proposed model. $\mu_h$, $\phi_h$ and $\sigma_h$ represent the log-scale level, persistence and standard deviation, respectively.$ \sigma_{\lambda}$ captures the dynamic of the skewness component and $\alpha_{0}$ is the initial level of the skewness. Both series are likely to have a dynamic skewness as show by the posterior summaries of $\sigma_{\lambda}$ with the Brazilian bonds having a bigger range of skewness variation than bonds from the US}
\label{Tab:ParametersProposal}
\end{table}

\begin{figure}[h!]
\centering
\begin{subfigure}[b]{0.95\textwidth}
\includegraphics[height=5cm,width=0.95\textwidth]{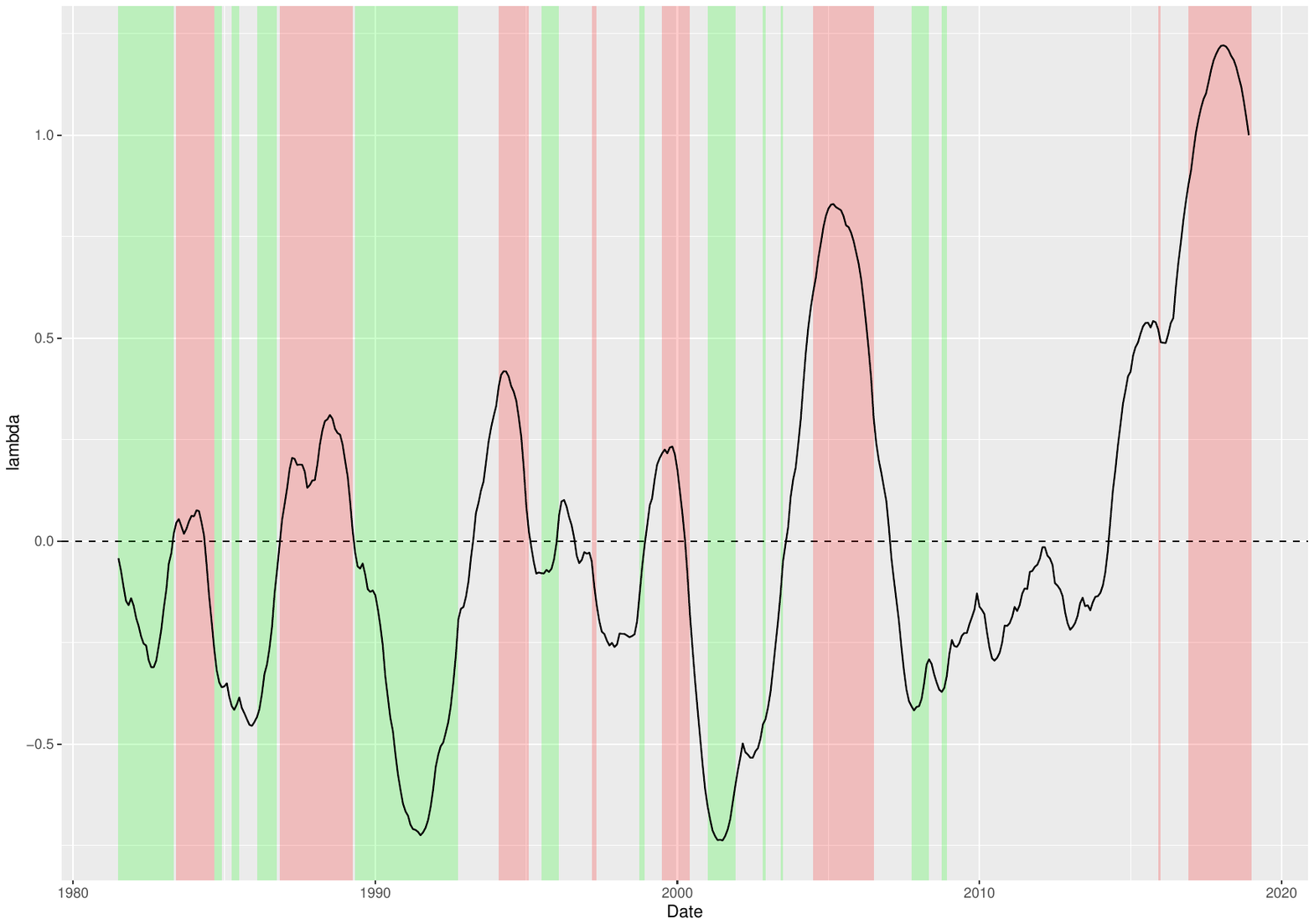} 
   \caption{}
   \label{fig:Us_lambda_cycle} 
\end{subfigure}

\begin{subfigure}[b]{0.95\textwidth}
\includegraphics[height=5cm,width=0.95\textwidth]{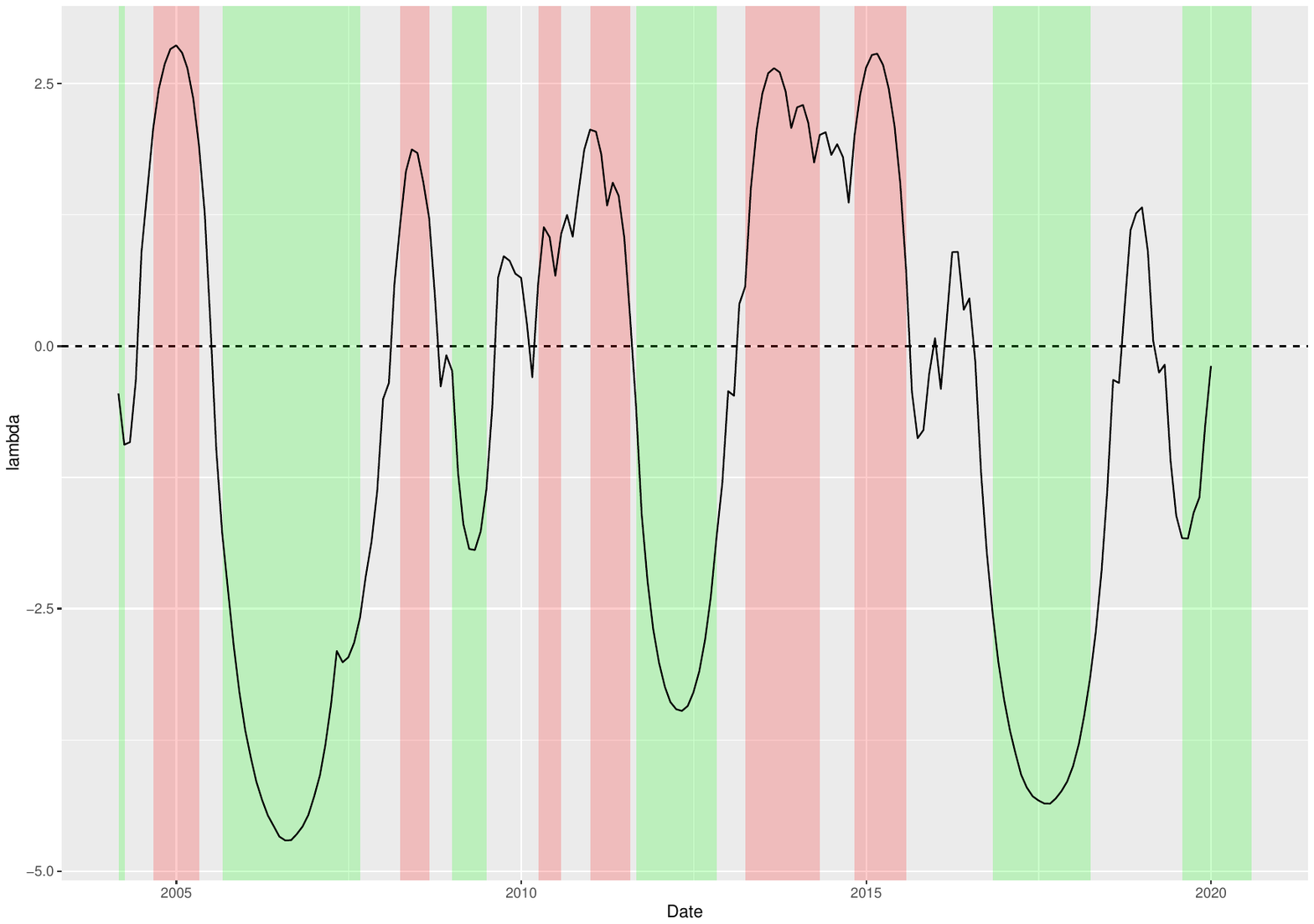} 
   \caption{}
   \label{fig:Br_lambda_cycle}
\end{subfigure}
\caption{$\{\lambda_t\}$ recovered for the US, top panel, and Brazilian bonds, bottom panel, during the estimation period which goes up to 12/2018 for the US and up to 12/2019 for the Brazilian case. Green shaded area indicate monetary easing periods while red ones indicate monetary tightening. In both cases, $\lambda_t$ seems able to capture the monetary cycles and reflect future direction of yield change. US monetary police cycles are based on the FED effective fund rate while for the Brazilian case they are recovered from the target rate policy rate decision of each meeting.}
\label{Fig:lambda_cycles}
\end{figure}

Figure (\ref{Fig:lambda_cycles}) plots the posterior mean of $\{\lambda_t\}$ alongside with monetary easing-tightening cycles. Green shaded areas represent monetary easing periods characterized by interest rates cuts by the countries' central bank. Conversely, red shaded areas represent monetary tightening periods characterized by interest rates hikes. Our approach identifies negative skewness in the yield changes during easing and large positive skewness when a central bank is hiking interest rates. US monetary police cycles are based on the FED effective fund rate while for the Brazilian case they are recovered from the target rate policy rate decision of each meeting. \footnote{Brazilian target policy rate is available at https://www.bcb.gov.br/controleinflacao/historicotaxasjuros}

If the asymmetry of yield changes is connected to interest rate cycles, then drivers of the monetary policy should help explain skewness. We test this hypothesis by regressing the posterior mean of $\{\lambda_t\}$, denoted $\hat{\lambda}$, into inflation and unemployment as represented by Equation (\ref{Eq:regression_lambda_cb}). Both variables are reasonable  ex-ante since central banks have mandates of price stability and full employment. 

\begin{equation}\label{Eq:regression_lambda_cb}
    \hat{\lambda}_t \sim N(X \beta, \sigma^2) \text{ with } X = [1, Inflation_t, Unemployment_t]
\end{equation}

Table (\ref{Tab:Yields_linear_reg}) indicates that unemployment levels partially explains the asymmetry in yield changes for both countries. The negative sign of $\beta_{unemployment}$ is reasonable since central banks usually fights high unemployment levels with interest rate cuts to promote consumption and, as show in Figure (\ref{Fig:lambda_cycles}), easing cycles are associated with negative skewness on yield changes. Additionally, for the Brazilian case, inflation is also likely to play a role. The sign of $\beta_{inflation}$ is also reasonable since central banks combat surges in inflation with interest rake hikes which, as show in Figure (\ref{Fig:lambda_cycles}), is associated with high values of skewness. Finally, while inflation was a problem in the US in the late 70's and early 80's, for the majority of our estimation sample the American economy didn't face inflationary pressures. Since such pressures didn't exist, the yields won't react to them. Therefore, the near 0 effect of $\beta_{\lambda}$ is justifiable. 

\begin{table}[h!]
\centering
\begin{tabular}{lccc}
  \hline
 & Intercept & $\beta_{Inflation}$ & $\beta_{Unemployment}$ \\ 
 \hline 
US q05 & 0.39 & -0.01 & -0.11 \\ 
US mean & 0.58 & 0.00 & -0.09 \\ 
US q95  & 0.73 & 0.01 & -0.06 \\ 
  \hline
BR q05 & -0.72 & 1.16 & -0.35 \\ 
BR mean & 0.39 & 2.47 & -0.23 \\ 
BR q95 & 1.57 & 3.52 & -0.09 \\ 
   \hline
\end{tabular}
\caption{Linear regression of the posterior mean of $\{ \lambda_t \}$ into inflation and unemployment during the estimation sample. For both countries unemployment partially explains the skewness in yield changes while the asymmetry in the Brazilian case is also partially explained by inflation.}
\label{Tab:Yields_linear_reg}
\end{table}

We also evaluate the out of sample performance of our proposed model. If skewness is informative about the likely direction of changes in bond yields, then $sign(E_t  \hat{\lambda}_{t+1})$ should agree with $sign(y_{t+1})$. We verify this claim within a increasing window framework where the first window corresponds to the estimation sample. For each window, we run our HMC procedure and obtain $sign(E_t \hat{\lambda}_{t+1})$. As shown in Table (\ref{Tab:oos_yields}), our procedure indicates For US bonds, $sign(y_{t+1})$ is equal to $sign(E_t\lambda_{t+1})$ 66.1\% of cases with average value of change being 20.6\% when correctly forecasted and only 8.7\% when wrong. Similarly, for the Brazilian case, $sign(y_{t+1})$ is equal to $sign(E_t[\lambda_{t+1}])$ in 72.7\% of cases. Additionally, the average magnitude of correctly forecasted yield changes is 8.2\% and only 2.4\% when wrongly predicted. Thus, the out of sample analysis support our claim of dynamic skewness in bond yield changes being a forecasting variable of future yield changes. 

\begin{table}[h]
\centering
\begin{tabular}{lcc}
\hline
                    & US (\%) & Brazil (\%) \\
\hline
Hit Ratio           & 66.1    & 72.7        \\
Avg when right      & 20.6    & 8.2         \\
Avg when wrong      & 8.7     & 2.4         \\
$y_{t+1} > 0$       & 44.6    & 54.5     \\
\hline
\end{tabular}
\caption{If bond yield skewness captures the likely direction of yield changes, then we should expected yield increases when $E_t[\lambda_{t+1]}>0 $ and, otherwise, decreases. Our model provides evidence supporting such claims. In both cases, our model correctly predicts the correct yield change direction in at least 66.1\% of times.}
\label{Tab:oos_yields}
\end{table}

\section{Empirical application: Carry factor}\label{Sec:Carry}

\begin{figure}[h!]
\centering
\includegraphics[height= 9cm,width=0.95\textwidth]{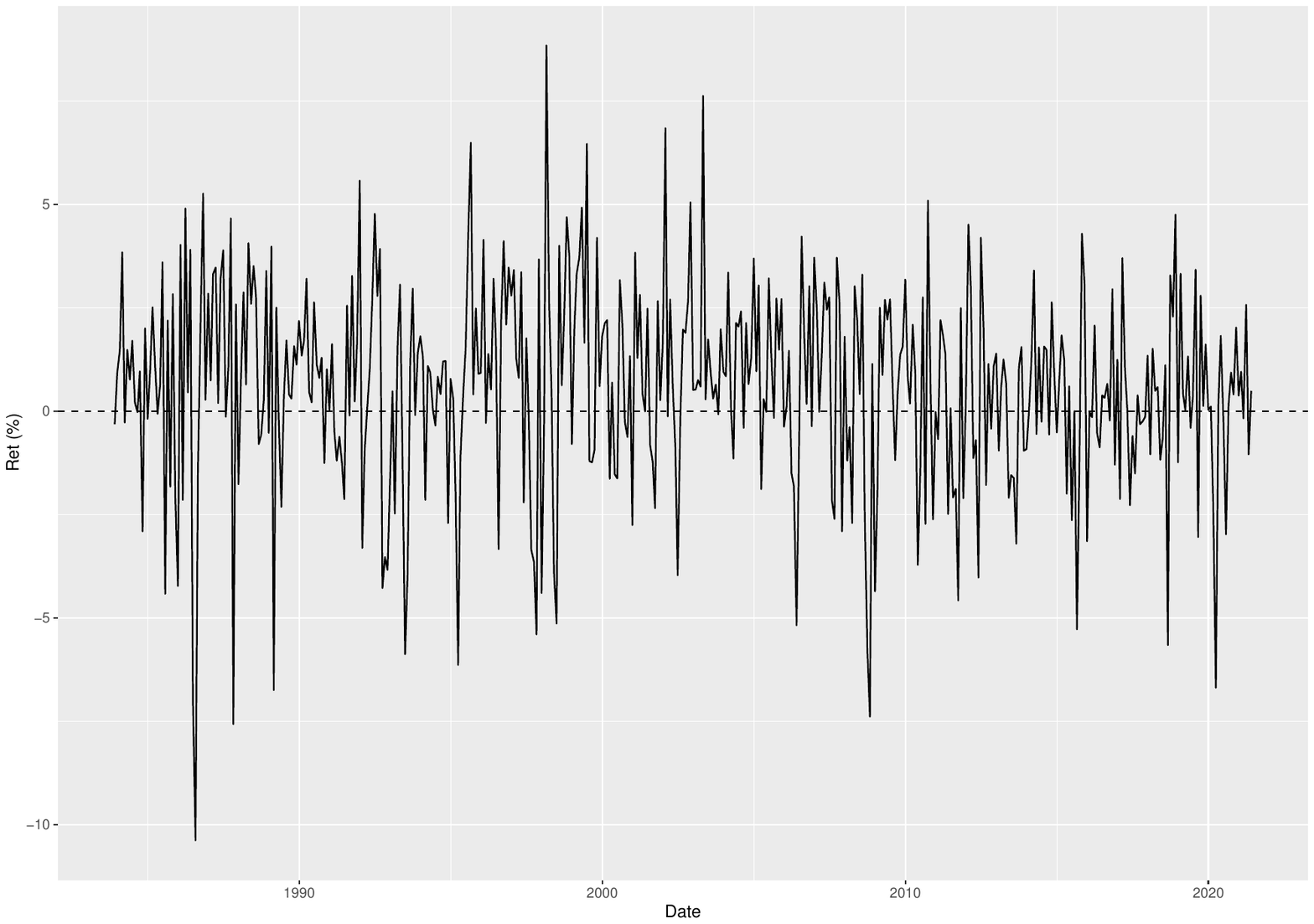}
\caption{Updated version of the time series for the Carry factor introduced by \cite{lustig2011common}. Our sample goes from 11/1983 to 05/2021.} 
\label{Fig:Carry_time_series}
\end{figure}

In addition to the Bond yield application described in Section \ref{Sec:BondYields}, we also consider an application for the  currency market. \cite{lustig2011common} indicates that two factors, dollar factor and carry, explain the cross-section of currency returns. This application focus on the latter which captures interest rate differentials between countries by going long on countries with high interest rate differentials with respect to the US and, conversely,  goes short countries with the smallest interest rate differentials with respect to the US. While we focus on FX markets, \cite{koijen2018carry} argues in favor of carry-based factors being a suitable to explain a the cross-section of a large number of asset classes such as commodities and equities. We consider and updated sample of the carry factor describe in \cite{lustig2011common}\footnote{You can check Lustig's carry factor on gsb-faculty.stanford.edu/hanno-lustig/files/2022/05/CurrencyPortfolios.xls}, presented in Figure (\ref{Fig:Carry_time_series}), which starts on 11/1983 and goes up to 05/2021.

We are not the first to study the skewness of carry returns. For example, \cite{burnside2011peso} and \cite{rafferty2012currency} identify a time-varying crash-risk on the carry factor. This risk would materialize in some occasions leading to large negative returns to the carry factor and skewing its distribution to the left. Conversely, \cite{jurek2014crash} uses out-of-the-money currency options hedging against large crashes and show that carry remains profitable indicating a small role for tail risk on currency returns. Additionally, \cite{jurek2014crash} presents evidence in favor of time varying volatility for carry returns and that the largest negative return of its sample, 10/2008, occurs in a period of high volatility. Our model is well suited to evaluate such claims. \cite{jurek2014crash} is not an isolated case. \cite{barroso2015momentum} provides another example of tail risk mitigation in tradeable factors after accounting for volatility without skewness playing a major role. 

\begin{figure}
\centering
\begin{subfigure}[b]{0.95\textwidth}
\includegraphics[height=7.5cm,width=0.95\textwidth]{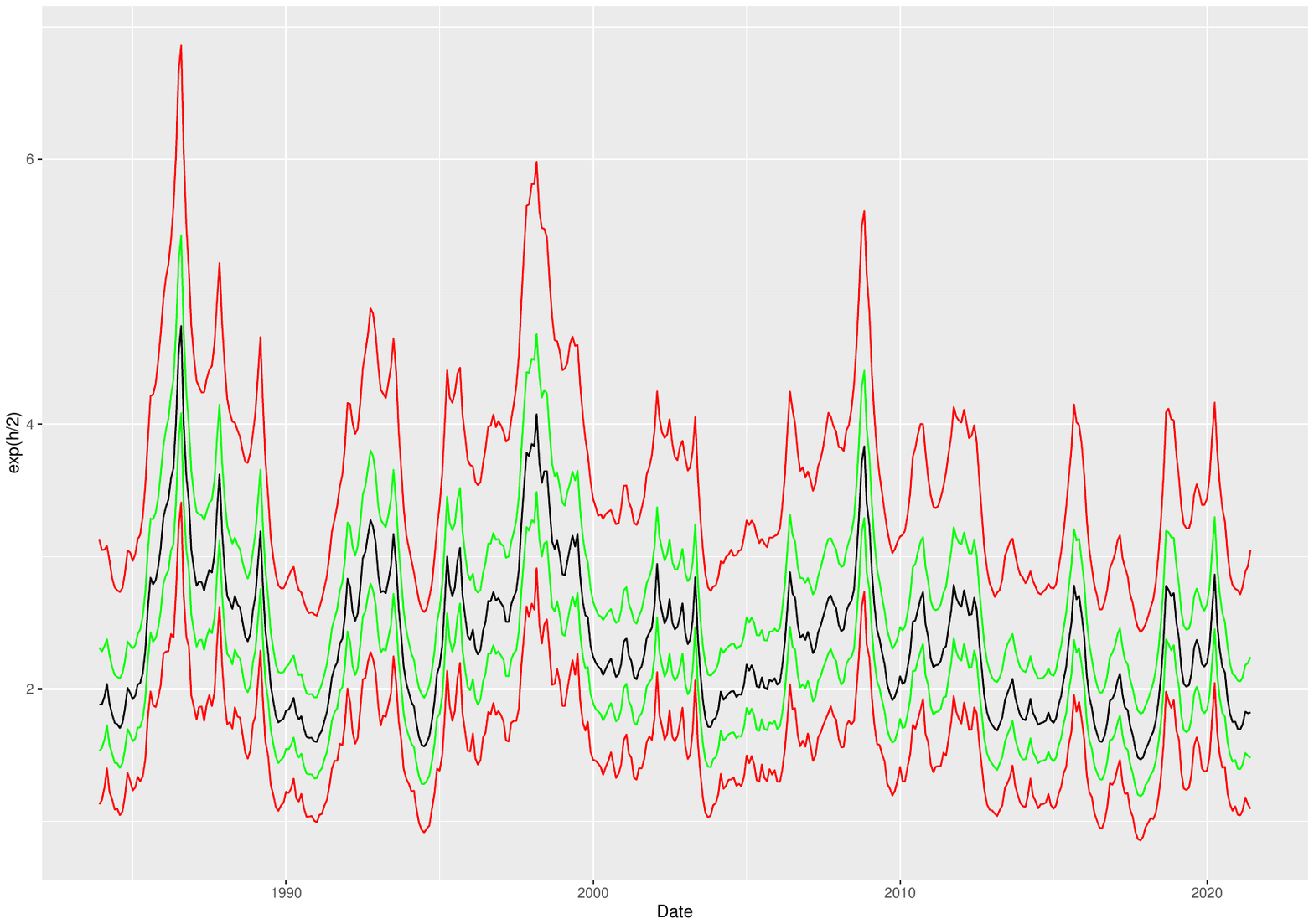} 
   \caption{}
   \label{fig:carry_exph2} 
\end{subfigure}

\begin{subfigure}[b]{0.95\textwidth}
\includegraphics[height=7.5cm,width=0.975\textwidth]{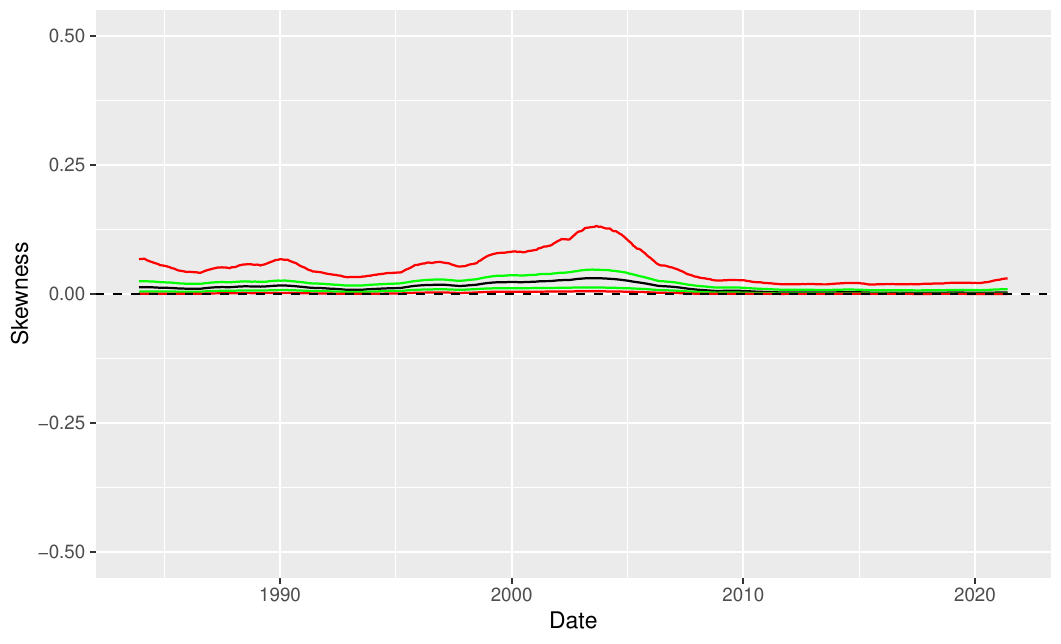} 
   \caption{}
   \label{fig:carry_skew}
\end{subfigure}
\caption{Skewness recovered from the carry factor. Green lines represent quantiles .25 and .75 while solid red lines represent quantiles .05 and .95. After controlling for volatility, we find no evidence of skewness in the carry factor.} 
\label{Fig:Carryexph2_skew}
\end{figure}

We consider the prior specification shown in Appendix C and sample from the posterior using the HMC scheme describe in Section \ref{Sec:Posterior}. Figure (\ref{Fig:Carryexph2_skew}) plots the posterior mean (black), interquartile range (green) and q05-q95 interval (red) for both $\{ exp \Big( \frac{h_t}{2} \Big)\}$ and $\{ \lambda_t \}$ which are shown at the top and bottom panel, respectively. The top panel corroborates with the evidence of time-varying volatility with a surge in volatility around 10/2008 similarly to the description of \cite{jurek2014crash}. The bottom panel indicate that is likely that there is no skewness at all after accounting for stochastic volatility. 

Additionally, we verify how the carry factor performs in period with high volatility when compared to low volatility. We say that volatility is high if is above the average volatility recovered for the full sample and we say its low otherwise. Table (\ref{Tab:Results_carry}) reports the results of our analysis. While the average return for the carry factor in both environments is the same, crash indicators such as the return on the 5th percentile and minimum return are severely improved. Therefore, in combination with the results shown in Figure (\ref{fig:carry_skew}), our results indicate that after accounting for volatility, its unlikely that skewness play a large role in affecting the returns of the carry factor.

\begin{table}[ht]
\centering
\begin{tabular}{lcc}
  \hline
 & High Vol & Low Vol \\ 
  \hline
Mean & 0.58 & 0.57 \\ 
Sd & 3.34 & 1.41 \\ 
Q05 & -5.38 & -2.07 \\ 
Min & -10.38 & -3.21 \\ 
   \hline
\end{tabular}
\caption{Summary statistics of the carry factor in high and low volatility environments for the sample starting in 11/1983 and going up to 12/2021. While the average return for the carry factor in both environments is the same, crash indicators such as the return on the 5-th percentile and minimum return are severely improved. }
\label{Tab:Results_carry}
\end{table}

\section{Conclusion} \label{Sec:Conclusion}

This paper expands stochastic volatility models by allowing for time-varying skewness without having to imposing it. By considering a LASSO-type regularization for both the standard deviation and starting level of the skewness dynamics, our model chooses between dynamic, static and no skewness in a data-driven approach. On our bond yield application, we highlight the benefits of dynamic skewness by showing its connection to monetary easing/tightening cycles and with central banks' mandates. Additionally, we show that asymmetry is informative about the likely direction of future bond yield changes. On the second application, we shed light into the debate of carry average returns reflecting time-varying skewness versus no skewness but time-varying volatility. Our model indicates no skewness after accounting for stochastic volatility.

\clearpage
\renewcommand\bibname{References}
\bibliography{biblio}

\begin{thebibliography}{36}
\providecommand{\natexlab}[1]{#1}
\providecommand{\url}[1]{\texttt{#1}}
\expandafter\ifx\csname urlstyle\endcsname\relax
  \providecommand{\doi}[1]{doi: #1}\else
  \providecommand{\doi}{doi: \begingroup \urlstyle{rm}\Url}\fi

\bibitem[Azzalini(1985)]{azzalini1985class}
A.~Azzalini.
\newblock A class of distributions which includes the normal ones.
\newblock \emph{Scandinavian journal of statistics}, pages 171--178, 1985.

\bibitem[Azzalini(2013)]{azzalini2013skew}
A.~Azzalini.
\newblock \emph{The skew-normal and related families}, volume~3.
\newblock Cambridge University Press, 2013.

\bibitem[Barroso and Santa-Clara(2015)]{barroso2015momentum}
P.~Barroso and P.~Santa-Clara.
\newblock Momentum has its moments.
\newblock \emph{Journal of Financial Economics}, 116\penalty0 (1):\penalty0
  111--120, 2015.

\bibitem[Bayes and Branco(2007)]{bayes2007bayesian}
C.~L. Bayes and M.~D. Branco.
\newblock Bayesian inference for the skewness parameter of the scalar
  skew-normal distribution.
\newblock \emph{Brazilian Journal of Probability and Statistics}, pages
  141--163, 2007.

\bibitem[Belmonte et~al.(2014)Belmonte, Koop, and
  Korobilis]{belmonte2014hierarchical}
M.~A. Belmonte, G.~Koop, and D.~Korobilis.
\newblock Hierarchical shrinkage in time-varying parameter models.
\newblock \emph{Journal of Forecasting}, 33\penalty0 (1):\penalty0 80--94,
  2014.

\bibitem[Betancourt(2017)]{betancourt2017conceptual}
M.~Betancourt.
\newblock A conceptual introduction to hamiltonian monte carlo.
\newblock \emph{arXiv preprint arXiv:1701.02434}, 2017.

\bibitem[Bianchi et~al.(2022)Bianchi, De~Polis, and
  Petrella]{bianchi2022taming}
D.~Bianchi, A.~De~Polis, and I.~Petrella.
\newblock Taming momentum crashes.
\newblock \emph{Available at SSRN 4182040}, 2022.

\bibitem[Bitto and Fr{\"u}hwirth-Schnatter(2019)]{bitto2019achieving}
A.~Bitto and S.~Fr{\"u}hwirth-Schnatter.
\newblock Achieving shrinkage in a time-varying parameter model framework.
\newblock \emph{Journal of Econometrics}, 210\penalty0 (1):\penalty0 75--97,
  2019.

\bibitem[Burnside et~al.(2011)Burnside, Eichenbaum, Kleshchelski, and
  Rebelo]{burnside2011peso}
C.~Burnside, M.~Eichenbaum, I.~Kleshchelski, and S.~Rebelo.
\newblock Do peso problems explain the returns to the carry trade?
\newblock \emph{The Review of Financial Studies}, 24\penalty0 (3):\penalty0
  853--891, 2011.

\bibitem[Carpenter et~al.(2017)Carpenter, Gelman, Hoffman, Lee, Goodrich,
  Betancourt, Brubaker, Guo, Li, and Riddell]{carpenter2017stan}
B.~Carpenter, A.~Gelman, M.~D. Hoffman, D.~Lee, B.~Goodrich, M.~Betancourt,
  M.~A. Brubaker, J.~Guo, P.~Li, and A.~Riddell.
\newblock Stan: A probabilistic programming language.
\newblock \emph{Journal of statistical software}, 76, 2017.

\bibitem[Carr and Wu(2007)]{carr2007stochastic}
P.~Carr and L.~Wu.
\newblock Stochastic skew in currency options.
\newblock \emph{Journal of Financial Economics}, 86\penalty0 (1):\penalty0
  213--247, 2007.

\bibitem[Cochrane and Piazzesi(2005)]{cochrane2005bond}
J.~H. Cochrane and M.~Piazzesi.
\newblock Bond risk premia.
\newblock \emph{American economic review}, 95\penalty0 (1):\penalty0 138--160,
  2005.

\bibitem[Collin-Dufresne and Goldstein(2002)]{collin2002bonds}
P.~Collin-Dufresne and R.~S. Goldstein.
\newblock Do bonds span the fixed income markets? theory and evidence for
  unspanned stochastic volatility.
\newblock \emph{The Journal of Finance}, 57\penalty0 (4):\penalty0 1685--1730,
  2002.

\bibitem[Gamerman and Lopes(2006)]{gamerman2006markov}
D.~Gamerman and H.~F. Lopes.
\newblock \emph{Markov chain Monte Carlo: stochastic simulation for Bayesian
  inference}.
\newblock CRC press, 2006.

\bibitem[George and McCulloch(1993)]{george1993variable}
E.~I. George and R.~E. McCulloch.
\newblock Variable selection via gibbs sampling.
\newblock \emph{Journal of the American Statistical Association}, 88\penalty0
  (423):\penalty0 881--889, 1993.

\bibitem[Guo et~al.(2020)Guo, Gabry, Goodrich, and Weber]{guo2020package}
J.~Guo, J.~Gabry, B.~Goodrich, and S.~Weber.
\newblock Package ‘rstan’.
\newblock \emph{URL https://cran.r―project.org/web/packages/rstan/}, 2020.

\bibitem[G{\"u}rkaynak et~al.(2007)G{\"u}rkaynak, Sack, and
  Wright]{gurkaynak2007us}
R.~S. G{\"u}rkaynak, B.~Sack, and J.~H. Wright.
\newblock The us treasury yield curve: 1961 to the present.
\newblock \emph{Journal of monetary Economics}, 54\penalty0 (8):\penalty0
  2291--2304, 2007.

\bibitem[Hansen and Richard(1987)]{hansen1987role}
L.~P. Hansen and S.~F. Richard.
\newblock The role of conditioning information in deducing testable
  restrictions implied by dynamic asset pricing models.
\newblock \emph{Econometrica: Journal of the Econometric Society}, pages
  587--613, 1987.

\bibitem[Jacquier et~al.(2002)Jacquier, Polson, and
  Rossi]{jacquier2002bayesian}
E.~Jacquier, N.~G. Polson, and P.~E. Rossi.
\newblock Bayesian analysis of stochastic volatility models.
\newblock \emph{Journal of Business \& Economic Statistics}, 20\penalty0
  (1):\penalty0 69--87, 2002.

\bibitem[Joslin(2018)]{joslin2018can}
S.~Joslin.
\newblock Can unspanned stochastic volatility models explain the cross section
  of bond volatilities?
\newblock \emph{Management Science}, 64\penalty0 (4):\penalty0 1707--1726,
  2018.

\bibitem[Jurek(2014)]{jurek2014crash}
J.~W. Jurek.
\newblock Crash-neutral currency carry trades.
\newblock \emph{Journal of Financial Economics}, 113\penalty0 (3):\penalty0
  325--347, 2014.

\bibitem[Koijen et~al.(2018)Koijen, Moskowitz, Pedersen, and
  Vrugt]{koijen2018carry}
R.~S. Koijen, T.~J. Moskowitz, L.~H. Pedersen, and E.~B. Vrugt.
\newblock Carry.
\newblock \emph{Journal of Financial Economics}, 127\penalty0 (2):\penalty0
  197--225, 2018.

\bibitem[Li and Scharth(2020)]{li2020leverage}
M.~Li and M.~Scharth.
\newblock Leverage, asymmetry and heavy tails in the high-dimensional factor
  stochastic volatility model.
\newblock \emph{Journal of Business \& Economic Statistics}, \penalty0
  (just-accepted):\penalty0 1--38, 2020.

\bibitem[Litterman and Scheinkman(1991)]{litterman1991common}
R.~B. Litterman and J.~Scheinkman.
\newblock Common factors affecting bond returns.
\newblock \emph{The journal of fixed income}, 1\penalty0 (1):\penalty0 54--61,
  1991.

\bibitem[Lopes et~al.(2022)Lopes, McCulloch, and Tsay]{lopes2022parsimony}
H.~F. Lopes, R.~E. McCulloch, and R.~S. Tsay.
\newblock Parsimony inducing priors for large scale state--space models.
\newblock \emph{Journal of Econometrics}, 230\penalty0 (1):\penalty0 39--61,
  2022.

\bibitem[Lustig et~al.(2011)Lustig, Roussanov, and Verdelhan]{lustig2011common}
H.~Lustig, N.~Roussanov, and A.~Verdelhan.
\newblock Common risk factors in currency markets.
\newblock \emph{The Review of Financial Studies}, 24\penalty0 (11):\penalty0
  3731--3777, 2011.

\bibitem[Markowitz(1952)]{markowitz1952}
H.~Markowitz.
\newblock Portfolio selection.
\newblock \emph{Journal of Finance}, 7\penalty0 (1):\penalty0 77--91, 1952.

\bibitem[Nakajima(2020)]{nakajima2020skew}
J.~Nakajima.
\newblock Skew selection for factor stochastic volatility models.
\newblock \emph{Journal of Applied Statistics}, 47\penalty0 (4):\penalty0
  582--601, 2020.

\bibitem[Nakajima and Omori(2012)]{nakajima2012stochastic}
J.~Nakajima and Y.~Omori.
\newblock Stochastic volatility model with leverage and asymmetrically
  heavy-tailed error using gh skew student’s t-distribution.
\newblock \emph{Computational Statistics \& Data Analysis}, 56\penalty0
  (11):\penalty0 3690--3704, 2012.

\bibitem[Park and Casella(2008)]{park2008bayesian}
T.~Park and G.~Casella.
\newblock The bayesian lasso.
\newblock \emph{Journal of the American Statistical Association}, 103\penalty0
  (482):\penalty0 681--686, 2008.

\bibitem[Rachev et~al.(2005)Rachev, Menn, and Fabozzi]{rachev2005fat}
S.~T. Rachev, C.~Menn, and F.~J. Fabozzi.
\newblock \emph{Fat-tailed and skewed asset return distributions: implications
  for risk management, portfolio selection, and option pricing}.
\newblock John Wiley \& Sons, 2005.

\bibitem[Rafferty(2012)]{rafferty2012currency}
B.~Rafferty.
\newblock Currency returns, skewness and crash risk.
\newblock \emph{Skewness and Crash Risk (March 15, 2012)}, 2012.

\bibitem[Shephard(2005)]{shephard2005stochastic}
N.~Shephard.
\newblock \emph{Stochastic volatility: selected readings}.
\newblock OUP Oxford, 2005.

\bibitem[Thomas and Tu(2021)]{thomas2021learning}
S.~Thomas and W.~Tu.
\newblock Learning hamiltonian monte carlo in r.
\newblock \emph{The American Statistician}, 75\penalty0 (4):\penalty0 403--413,
  2021.

\bibitem[Tibshirani(1996)]{tibshirani1996regression}
R.~Tibshirani.
\newblock Regression shrinkage and selection via the lasso.
\newblock \emph{Journal of the Royal Statistical Society: Series B
  (Methodological)}, 58\penalty0 (1):\penalty0 267--288, 1996.

\bibitem[Trolle and Schwartz(2014)]{trolle2014swaption}
A.~B. Trolle and E.~S. Schwartz.
\newblock The swaption cube.
\newblock \emph{The Review of Financial Studies}, 27\penalty0 (8):\penalty0
  2307--2353, 2014.

\end{thebibliography}

\clearpage
\section*{Appendix A: Moments of a skew-normal distribution}

Following \cite{azzalini1985class} and \cite{bayes2007bayesian},  $\xi$, $\omega$ and $\lambda$ relate to mean, variance and skewness by the following expressions:

If $Z \sim SN(\xi, \omega, \lambda)$, then 

$$ E[Z] = \xi + \sqrt{\omega}\delta \sqrt{\frac{2}{\pi}} $$

$$ Var[Z] = \omega \Bigg( 1 - \frac{2}{\pi} \delta^2 \Bigg)$$

$$ \gamma = \sqrt{\frac{2}{\pi}} \Bigg( \frac{4}{\pi} - 1 \Bigg) \Bigg( \frac{\lambda}{\sqrt{1 + \lambda^2}} \Bigg)^3 \Bigg( 1 - \frac{2}{\pi} \frac{\lambda^2}{1 + \lambda^2} \Bigg)^{3/2}$$
where $\delta = \frac{\lambda}{\sqrt{1 + \lambda^2}}$ and $\gamma$ is the skewness index. Note that $\gamma$ depends only $\lambda$. 

\clearpage
\section*{Appendix B: Priors and additional results for the Bond applications}

We consider the following structure for the bond yield applications 

$$ \mu_h \sim N(4,10^2)$$
$$ \phi_h \sim N(0.95,0.5^2)$$
$$ \sigma \sim IG(0.1,0.1) $$
$$ \alpha_0 \sim DoubleExp(0,1/\kappa_{\alpha})$$
$$ \sigma_{\lambda} \sim DoubleExp(0,1/\kappa_{\sigma_{\lambda}})$$
$$ \kappa_{\alpha} \sim Gamma(0.1, 0.1) $$
$$ \kappa_{\sigma_{\lambda}}\sim Gamma(0.1, 0.1) $$

We also plot the skewness itself using green and red shades to indicate easing and tightening periods in addition to the posterior for $\{\lambda_t\}$. While the conclusions are the same, the version with skewness may help the reader to make a better assessment of the magnitude of skewness in each period.

\begin{figure}[h!]
\centering
\begin{subfigure}[b]{0.95\textwidth}
\includegraphics[height=7cm,width=0.95\textwidth]{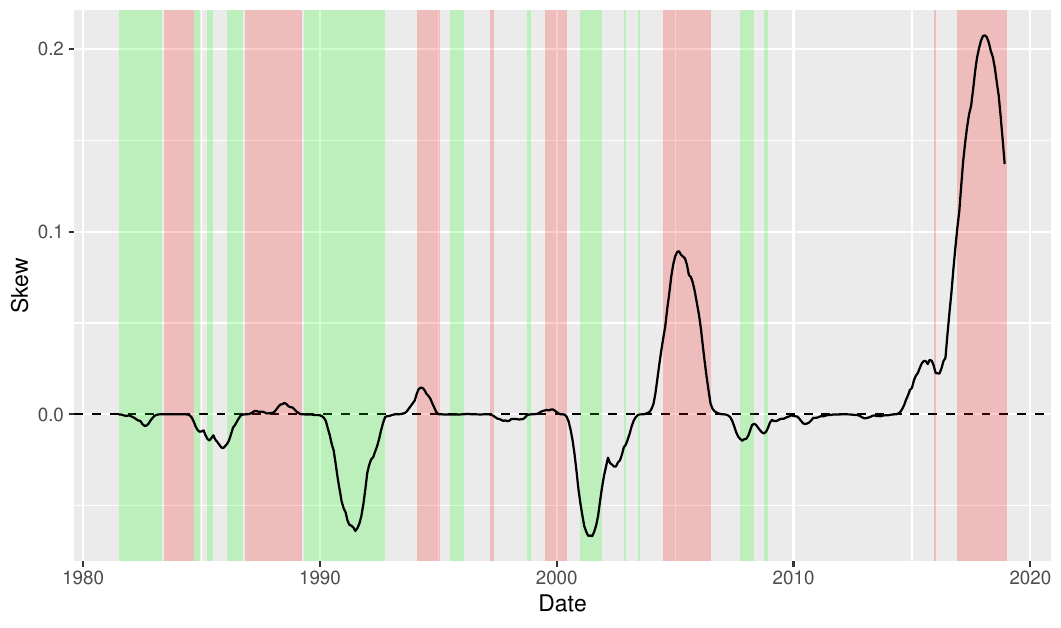} 
   \caption{}
   \label{fig:us_skew_cycle} 
\end{subfigure}

\begin{subfigure}[b]{0.95\textwidth}
\includegraphics[height=7cm,width=0.95\textwidth]{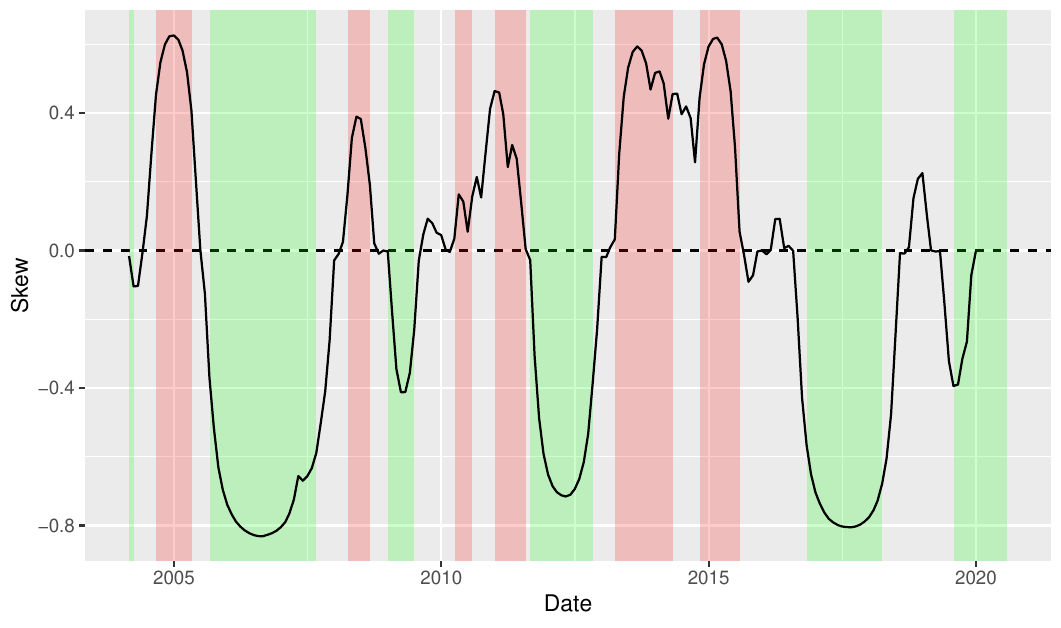} 
   \caption{}
   \label{fig:br_skew_cycle}
\end{subfigure}
\caption{Skewness recovered based on the posterior median of $\{\lambda_t\}$ for the US, top panel, and Brazilian bonds, bottom panel, during the estimation period which goes up to 12/2018 for the US and up to 12/2019 for the Brazilian case. Green shaded area indicate monetary easing periods while red ones indicate monetary tightening. US monetary police cycles are based on the FED funds effective rate changes while for the Brazilian case they are recovered from the target rate policy rate decision of each meeting. In both countries, positive skewness is associated with tight monetary policy periods while negative skewness is connected to easing cycles. }
\end{figure}

\clearpage
\section*{Appendix C: Priors for the FX application}

We consider the following structure for the bond yield applications 

$$ \mu_h \sim N(4,10^2)$$
$$ \phi_h \sim N(0.95,0.5^2)$$
$$ \sigma_h \sim IG(2,2) $$
$$ \alpha_0 \sim DoubleExp(0,1/\kappa_{\alpha})$$
$$ \sigma_{\lambda} \sim DoubleExp(0,1/\kappa_{\sigma_{\lambda}})$$
$$ \kappa_{\alpha} \sim Gamma(0.1, 0.1) $$
$$ \kappa_{\sigma_{\lambda}}\sim Gamma(0.1, 0.1) $$

\end{document}